\documentclass[letterpaper,twocolumn,10pt]{article}
\usepackage{hyperref}
\usepackage{usenix2019_v3}

\usepackage[linesnumbered,ruled]{algorithm2e}

\usepackage{mathtools}
\usepackage[font=footnotesize,labelfont=bf]{caption}
\usepackage{booktabs}
\usepackage{subcaption}
\usepackage[scaled=0.8]{beramono} %
\usepackage{listings}

\usepackage{titling}
\definecolor{mygreen}{rgb}{0.1, 0.5, 0.1}

\usepackage{newunicodechar}
\newunicodechar{λ}{\ensuremath{\lambda}}
\newunicodechar{≈}{\ensuremath{\approx}}

\usepackage{enumitem}
\setitemize{noitemsep,topsep=0pt,parsep=0pt,partopsep=0pt}
\setlistdepth{5}
\setenumerate{noitemsep,topsep=0pt,parsep=0pt,partopsep=0pt}
\setdescription{itemsep=0pt,topsep=0pt,parsep=0pt,partopsep=0pt,itemindent=0em,leftmargin=0.5em}

\newenvironment{denseitemize}{
\begin{itemize}[topsep=2pt, partopsep=0pt, leftmargin=1.5em]
  \setlength{\itemsep}{2pt}
  \setlength{\parskip}{0pt}
  \setlength{\parsep}{0pt}
}{\end{itemize}}

\newenvironment{denseenum}{
\begin{enumerate}[topsep=2pt, partopsep=0pt, leftmargin=1.5em]
  \setlength{\itemsep}{2pt}
  \setlength{\parskip}{0pt}
  \setlength{\parsep}{0pt}
}{\end{enumerate}}

\def\ie{{i.e.\ }}
\def\eg{{e.g.,\ }}

\usepackage{authblk}
\makeatletter
\renewcommand\AB@affilsepx{, \protect\Affilfont}
\makeatother

\def\name{\texttt{cachebpf}\xspace}

\begin{document}
\date{\vspace{-1.3cm}}

\title{Cache is King: Smart Page Eviction with eBPF}
\author[*,1]{Tal Zussman}
\author[*,1]{Ioannis Zarkadas}
\author[1]{Jeremy Carin}
\author[1]{Andrew Cheng}
\author[2]{Hubertus Franke}
\author[2]{Jonas Pfefferle}
\author[1]{Asaf Cidon}
\affil[1]{Columbia University}
\affil[2]{IBM}
\affil[*]{denotes equal contribution}

\pagestyle{plain}

\maketitle

\begin{abstract}

The page cache is a central part of an OS. It reduces repeated accesses to storage by deciding which pages to retain in memory. As a result, the page cache has a significant impact on the performance of many applications. However, its one-size-fits-all eviction policy performs poorly in many workloads. While the systems community has experimented with a plethora of new and adaptive eviction policies in non-OS settings (\eg key-value stores, CDNs), it is very difficult to implement such policies in the page cache, due to the complexity of modifying kernel code. To address these shortcomings, we design a novel eBPF-based framework for the Linux page cache, called \name, that allows developers to customize the page cache without modifying the kernel.
\name enables applications to customize the page cache policy for their specific needs, while also ensuring that different applications' policies do not interfere with each other and preserving the page cache's ability to share memory across different processes.
We demonstrate the flexibility of \name's interface by using it to implement several eviction policies. Our evaluation shows that it is indeed beneficial for applications to customize the page cache to match their workloads' unique properties, and that they can achieve up to 70\% higher throughput and 58\% lower tail latency.

\end{abstract}

\section{Introduction}

In his seminal 1981 paper on OS support for database management, Michael Stonebraker described how existing OS buffer cache mechanisms were ill-suited for the needs of databases at the time~\cite{stonebraker_support}. He observed that the buffer cache's one-size-fits-all eviction policy, approximate least-recently used (LRU), cannot possibly address the heterogeneity of database workloads. Nevertheless, in the intervening decades, despite wide-ranging efforts to rethink the UNIX/Linux OS page cache~\cite{parallel-page-cache,mglru,new-readahead}, design customizable file systems~\cite{kai-li-caching,kai-li-caching2,exokernel, bento}, and build clean-slate extensible kernels~\cite{VINO,VINO-2,SPIN}, applications by and large still contend with Linux's opaque and inflexible OS page cache policy. 

At the same time, the diversity of applications and workloads running on Linux has only increased, from enterprise file systems and large-scale distributed datacenter ML training, to multimedia rich applications running on an Android phone. All of these applications must use Linux's decades-old approximate LRU policy, despite the fact it is widely known to be inadequate for many workloads and scenarios (e.g., large scans~\cite{LIRS,frequency-replacement,LRU-K}, multi-core applications~\cite{parallel-page-cache}).
For example, an application that searches through files in a codebase (a scan-based workload) would benefit from using a most-recently used (MRU) policy, while a key-value store running a fixed, skewed-distribution workload would improve under least-frequently used (LFU). However, both of these workloads currently run with the default Linux eviction policy.

The reasons applications are ``stuck'' with the same old eviction policy are twofold. First, modifying the Linux page cache is a hard task, requiring extensive kernel knowledge and attention to detail. Second, upstreaming changes to the page cache is difficult, %
because the changes have to work well for the wide range of applications that run on Linux, forcing a lowest common denominator. For instance, it took Google years to upstream its proposed Multi-Generational LRU (MGLRU) algorithm into the Linux kernel, and even after several years, it is still disabled by default in upstream~\cite{mglru,mglru2}.

In this paper, we attempt to finally answer Stonebraker's plea for better OS support for buffer management, within Linux. To this end, we design a novel framework, \name, which provides visibility and control of the OS page cache, without requiring the application to make kernel changes. \name takes advantage of eBPF~\cite{ebpf}, a Linux (and Windows) supported runtime that allows safely running application code inside the kernel. We take a cue from \texttt{sched\_ext}, an eBPF-based framework that allows applications to customize the OS scheduler~\cite{syrup,ghost} and has been adopted by Linux~\cite{sched_ext}.

\name's design is motivated by four main insights. First, modern storage devices are very fast and support millions of IOPS, so custom page cache policies must run with low overhead. Therefore, we design \name so that its eBPF-based policies run in the kernel, avoiding expensive and frequent synchronization between the kernel and userspace.
Second, caching algorithms are very diverse and may use complex data structures. %
To address this challenge, \name exposes a simple yet flexible interface that allows applications to define one or more variable-sized lists of pages, and a set of policy functions (\eg admission, eviction) that operate on these lists, which can be used to express a wide range of eviction policies.
Third, in order for \name to be useful in multi-tenant scenarios, it should allow each application to use its own policy without interfering with others. We identify cgroups as a natural isolation boundary. Thus, \name allows each cgroup to implement its own eviction policy without interfering with other cgroups.
Finally, custom policies determine which pages to evict and return page references to the kernel. However, these references may be invalid, which could lead to kernel crashes or security breaches. To solve this, \name maintains a registry of valid page references, which is used to validate the page references returned by the user-defined policies.

We demonstrate \name's utility and flexibility by implementing four custom eviction policies, which include both ``classic'' and recently-designed policies: most-recently used (MRU), least-frequently used (LFU), S3-FIFO~\cite{s3-fifo}, and least hit density (LHD)~\cite{lhd}.
We also show how \name enables \emph{application-informed} policies with only minor policy changes, allowing applications to design policies that take into account application-level insights.
For example, a database can implement a custom policy that prioritizes point queries over scans, yielding higher throughput for point queries.
We compare these \name policies with the kernel's default eviction policy and its different options (\eg \texttt{fadvise()}), and with the recently-upstreamed MGLRU algorithm. We show that with \name, developers can significantly improve their applications' performance far beyond the existing algorithms provided by the Linux page cache.
In general, we find that there is no one-size-fits-all policy that improves all workloads -- customization is necessary in order to maximize performance. %
In particular, applications can use \name to improve throughput by up to 38\% using ``generic'' policies, and achieve up to 70\% higher throughput and 58\% lower P99 latency with application-informed policies.

We will open source \name and all our implemented policies upon publication. A key benefit of \name is that any publicly available eviction policy can be used by other developers, lowering the barrier to using the system and experimenting with eviction policies on different workloads. 

Our primary contributions are:
\begin{denseitemize}
\item \name, a flexible, scalable, and safe eBPF-based framework for running custom eviction policies in the Linux kernel page cache.
\item A suite of custom eviction policies and userspace libraries allowing developers to easily create new policies.
\item An evaluation of \name across various applications, demonstrating how they benefit from customized policies.
\end{denseitemize}

\section{Background and Motivation}
\label{sec:background}

By default, the page cache buffers write and read operations to and from storage devices. %
In Linux, the page cache tracks pages and stores them in lists (see \S\ref{sec:primer}), on which it approximates the LRU algorithm.
While this scheme works reasonably well for some workloads, it is inadequate for many others. The classic example is scan-heavy workloads, which perform poorly with LRU or its approximations~\cite{lhd,stonebraker_support,arc}. While Linux provides some interfaces (\eg \texttt{fadvise()} or \texttt{sysctl}) through which the page cache behavior can be tweaked on a global or per-application basis, these interfaces are opaque and do not perform as intended, as we describe in \S\ref{sec:primer} and evaluate in \S\ref{sec:app-informed}.

Therefore, to avoid compromising performance, some applications implement their own userspace-based caches~\cite{memory-rocksdb,wiredtiger,postgres-cache,innodb}. However, userspace-based caches are not a panacea. First, they require significant developer effort to implement. Second, they typically require the application to specify in advance how much memory will be allocated for the cache. However, the amount of memory available to an application may change over time (\eg when multiple applications run on the same physical server). Third, application-specific caches are very hard to share across processes, due to security and compatibility issues. Ultimately, even applications that implement their own userspace-based cache often still rely on the page cache by default as a ``second-tier'' cache~\cite{memory-rocksdb,wiredtiger,postgres-cache}, allowing operators to fully utilize the server's memory and share memory across processes.
As such, despite the page cache's limitations, it is still used extensively by storage-optimized workloads, such as key-value stores~\cite{rocksdb,leveldb,wiredtiger}, databases~\cite{postgres,innodb}, and ML inference and training systems~\cite{pytorch-load,milvus-load}.

Unfortunately, these factors yield a status quo where potential performance gains are left on the table. Properly customizing the page cache is not an easy task -- it is deeply intertwined with other performance- and correctness-critical memory management and filesystem code paths. While work to modernize the page cache is ongoing, it does not yet seem to have achieved this goal. In particular, MGLRU, an alternative LRU implementation for the page cache, has still not been enabled by default in upstream Linux several years after it was introduced, %
and it does not provide customization interfaces~\cite{mglru,mglru2}. %
Indeed, in \S\ref{sec:evaluation} we show that MGLRU sometimes underperforms and sometimes outperforms the default LRU algorithm, and that in general there is no single eviction policy that performs best across a wide range of workloads.

We now provide a primer on the Linux page cache. We also describe the eBPF framework, which \name uses to allow applications to write custom page cache policies.

\subsection{Linux Page Cache}
\label{sec:primer}

\begin{figure}[ht]
    \centering
    \includegraphics[width=0.7\columnwidth]{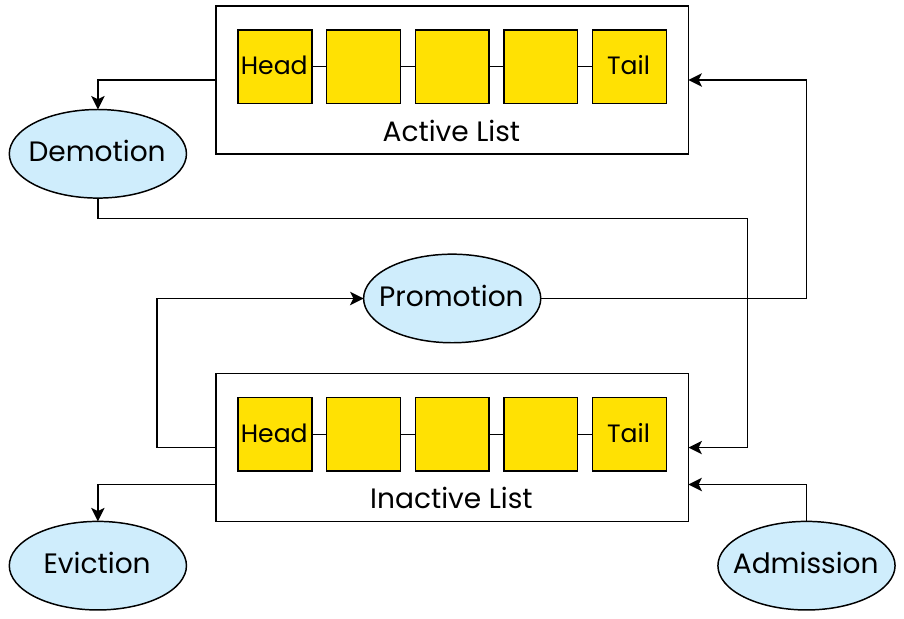}
    \caption{Overview of the current Linux page cache eviction policy.}
    \label{fig:default-page-cache-policy}
\end{figure}

The page cache is a core component of the Linux kernel, responsible for accelerating access to storage. While anonymous memory pages are stored similarly to file-backed memory, in this paper we focus specifically on file-backed memory. %
The kernel's default eviction policy is an LRU approximation algorithm which uses two FIFO lists.\footnote{The Linux page cache algorithm description is based on Linux v6.6.8.} As shown in Figure~\ref{fig:default-page-cache-policy}, when a page is first fetched from storage, it is added to the tail of the inactive list. If that page is accessed again, it will eventually be promoted to the active list. The goal of this policy is to use the inactive list as a preliminary filter and keep frequently accessed pages in the active list. When eviction is triggered, pages are removed from the head of the inactive list. If necessary, the page cache will balance the lists by demoting pages from the head of the active list to the tail of the inactive list. Notably, during balancing or shrinking, pages in the active list that have been referenced are typically demoted to the inactive list, rather than being given another chance in the active list, as is typical for LRU or CLOCK algorithms.

Importantly, active and inactive lists are segmented by cgroup. cgroups are a Linux feature which isolate resource usage for groups of processes~\cite{cgroup-v2}. Each cgroup has its own set of page cache lists which count toward its memory allocation, allowing for cgroup-specific eviction when its memory threshold is reached. Processes in cgroup A can access a page ``owned'' by cgroup B -- such an access will update the page's metadata (affecting its placement in cgroup B's lists), but will not count against cgroup A's memory limit. The combination of these per-cgroup lists make up the page cache as a whole.\footnote{Technically, each NUMA node has its own set of per-cgroup lists, but this does not affect our design.}

The page cache also keeps track of ``shadow entries'' in order to mitigate thrashing. These entries keep track of metadata enabling calculation of a page's refault distance (\ie the time elapsed between eviction and the new request). If a page has been evicted and then fetched again recently enough, the kernel may decide to insert it directly into the active list instead of the inactive list. There are several additional edge cases and heuristics in the kernel's implementation, but these are the broad strokes of the existing policy.

\vspace{-1em}
\paragraph{Folios.} Linux developers are in the process of converting various usages of \texttt{struct page} to \emph{folios}, which represent either zero-order pages (a single page) or the head page of a compound page (a group of contiguous physical pages that can be treated as a single larger page)~\cite{folios-lwn}. While the page cache now largely uses folios, we use the terms ``folio'' and ``page'' interchangeably, as in our workloads all folios represent a single page.

\vspace{-1em}
\paragraph{Userspace interfaces.} While LRU is a commonly-used eviction policy that works well across many workloads, there are many applications that would benefit from a different policy for part or the entirety of their I/O requests. For example, LRU is notoriously bad for scan-like access patterns. This gap between applications and the kernel can be partially mitigated by the \texttt{madvise()} and \texttt{fadvise()} system calls. These interfaces allow userspace applications to give \emph{hints} to the kernel about how to handle certain ranges of memory or files.

While these hints may help in simple cases, we show in our evaluation that they don't function as expected for some workloads. Additionally, while the hints may have a semantic meaning, their actual behavior is highly dependent on the kernel implementation, which is opaque, may change across versions, and can yield unexpected results~\cite{madvise-man-pages,madvise-surge-2015}.
Advice values may also be ignored by the kernel for a range of reasons, or may have restrictions on what memory they can be applied to.
Most importantly, these hints are still subject to the basic inflexible structure of the kernel's approximate LRU policy. 

\subsection{eBPF}

eBPF~\cite{ebpf} is a sandboxing technology that enables userspace functions to run in the Linux kernel in a safe and controlled manner. %
eBPF has found many use cases, including observability~\cite{gc-observability-bpf}, security~\cite{bpf-iptables,seccomp-bpf}, scheduling~\cite{sched_ext, syrup, ghost}, and I/O acceleration~\cite{xrp, xdp, electrode, dint, bpfof}. eBPF programs are \emph{verified} by the kernel before they can be run, ensuring, for example, that the programs don't contain illegal memory accesses, and that they will terminate within a fixed number of instructions. %

\section{Challenges}
\label{sec:challenges}

There are several challenges in allowing applications to customize the page cache using eBPF. We describe them below.

\begin{denseenum}
\item \textbf{Scalability.} Modern SSDs support millions of IOPS~\cite{solidigm-specs,western-digital-nvme-specs}, requiring the page cache to efficiently handle millions of events a second. Any changes to the page cache in order to enable custom policies must incur a low overhead, and the policies themselves must also be efficient.

\item \textbf{Flexibility.} Researchers have proposed many different caching algorithms for different use cases. These algorithms often require custom data structures. Any interface for custom policies must be flexible enough to accommodate the diversity of existing caching algorithms.

\item \textbf{Isolation and sharing.} The page cache is shared by many applications. Therefore, we must avoid a situation where one application's policy interferes with those of other applications, while still allowing applications to benefit from the shared nature of the page cache.

\item \textbf{Memory safety.} Custom eviction policies return page references to the kernel to indicate which pages to evict. This must not lead to unsafe memory references. %

\end{denseenum}

\section{Design and Implementation}
\label{sec:design}

\begin{figure}[t!]
    \centering
    \includegraphics[width=0.35\textwidth]{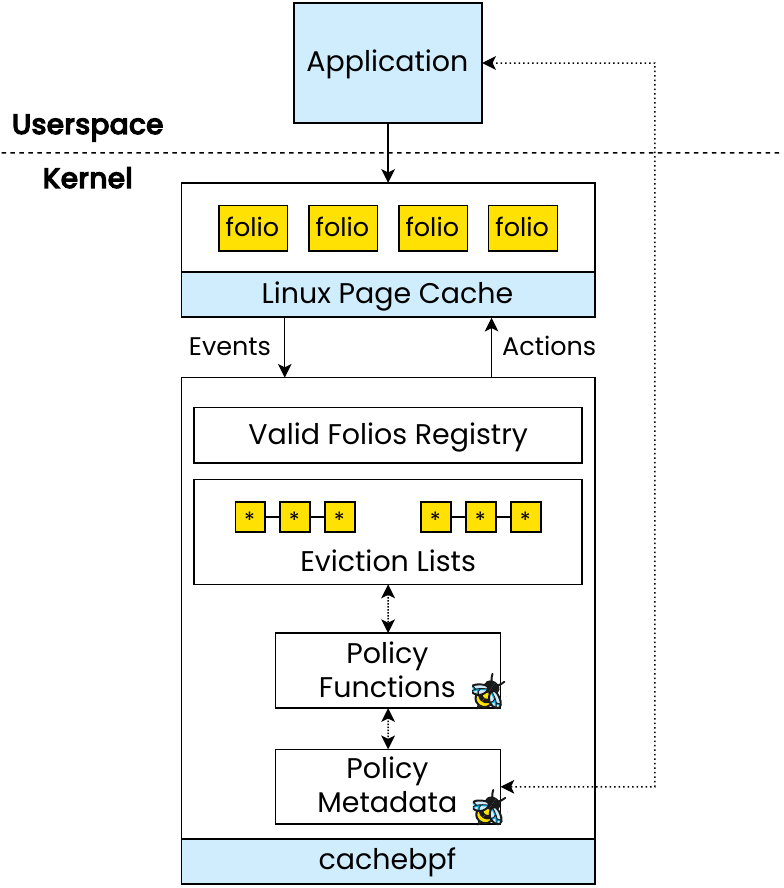}
    \caption{Overview of \name. Eviction lists hold pointers to folios.}
    \label{fig:system-diagram}
\end{figure}

In this section, we present \name's architecture and discuss how it addresses the challenges described in \S\ref{sec:challenges}.
Figure~\ref{fig:system-diagram} shows a diagram of the system.
At a high level, \name allows users to run custom eviction \emph{policy functions}, which are implemented as eBPF functions in the kernel. The policy functions are triggered by particular events (\eg folio eviction, access, admission), and they operate on a user-specified number of variable-sized \emph{eviction lists}, which store \emph{pointers} to the folios managed by the policy. The policy functions decide which folios to admit or evict to and from the lists based on metadata (\eg folio access frequency and recency, which thread accessed the folio), which is stored in eBPF maps. At eviction time, \name runs a user-defined eviction function to propose a set of \emph{eviction candidates} for the kernel to evict.
We find that while this interface is relatively simple, it is quite flexible, and can support a very wide set of eviction policies from the literature either exactly, or approximately.

We now describe \name's design in detail, starting with our design choice to implement \name's eviction policies within the kernel, rather than in userspace, to ensure scalability (challenge 1 from \S\ref{sec:challenges}).

\subsection{Policies in Kernel or Userspace?}
\label{subsec:kernel-vs-userspace}

Our first key design decision is whether to run \name's policies in the kernel or in userspace.
While from a development standpoint it might be simpler to run the eviction policies in userspace, doing so would require notifying userspace about all page cache events. However, modern SSDs can service millions of IOPS, each of which may trigger a page cache event (\eg folio access, insertion). 

We run a set of experiments to estimate the ``best-case'' overhead of such a userspace-offload architecture. We attach eBPF programs to existing kernel tracepoints (folio inserted, accessed, and evicted). The eBPF programs use a lockless ring buffer to notify userspace on each event~\cite{bpf-ringbuf}. Since no userspace logic actually processes these events, this provides an optimistic measure of this architecture's overhead.

We run two applications on a standard enterprise SSD to evaluate this architecture: YCSB workloads using RocksDB~\cite{rocksdb}, a key-value store, and a file search workload using ripgrep~\cite{ripgrep}, a parallelized grep-like tool, where we search the Linux kernel sources 10 times. The workloads are allocated 8~GiB and 1~GiB of memory, respectively. We run these applications on both the baseline system and with the eBPF benchmark programs. The results are presented in Table~\ref{tab:motivation-overhead}, with the benchmark yielding up to a 20.6\% performance decrease, \emph{without even implementing a custom eviction policy.} 

\begin{table}
    \centering
    \footnotesize
    \begin{tabular}{lrrr}
        \toprule
        Workload & Baseline    & Benchmark   & \% Degradation \\
        \midrule
        YCSB A   & 82,808 op/s & 69,089 op/s & -16.6\% \\
        YCSB C   & 76,166 op/s & 62,578 op/s & -17.8\% \\
        Uniform  & 44,618 op/s & 35,443 op/s & -20.6\% \\
        Search   &       42.3s &       44.4s &  -4.7\% \\
        \bottomrule
    \end{tabular}
    \caption{Performance of workloads without and with userspace-dispatch.}
    \label{tab:motivation-overhead}
\end{table}

Thus, based on the results of our experiments, we rule out implementing page cache policies in userspace, and instead opt to run the policies within the kernel as eBPF functions. We decide to use eBPF as it has already proven to match the kernel's performance, even in performance-critical domains such as networking~\cite{xdp} and storage~\cite{xrp}. While eBPF programs face many restrictions due to the verifier, we find that \name can provide sufficient flexibility for custom policies, as we describe below.

\subsection{Interface}
\label{subsec:kernel-impl}
Caching is an active area of research, with many recently-proposed eviction and admission algorithms~\cite{adaptsize,robinhood,lhd,lrb,baleen,sieve,s3-fifo,ripq} that aim to take advantage of different features of a workload (\eg recency, frequency, size), using various techniques (\eg conditional probability models~\cite{lhd}, Markov chains~\cite{adaptsize}, machine learning~\cite{lrb}).
To ensure flexibility, \name should allow developers to experiment with a wide range of caching policies, including relatively sophisticated ones. We now describe \name's API and demonstrate how it can be used to create a wide range of policies, addressing challenge 2 in \S\ref{sec:challenges}.

\subsubsection{Policy Functions}
\label{subsubsec:design_policy_functions}

\name allows applications to define custom eviction policies %
as \emph{policy functions}, a set of eBPF programs that trace caching events and determine which folios should be evicted from the page cache. 
Policy functions can be triggered by five events: policy initialization, requests for eviction, folio admission, folio access, and folio removal. The policy function interface is implemented using eBPF's recent \texttt{struct\_ops} kernel interface~\cite{struct-ops}, as shown in Figure~\ref{fig:cache-ext-ops}.

These five events are central to caching decisions in the page cache. %
Notably, requests for eviction and folio removal are different: the former involves the kernel asking the policy to propose folios to evict, and the latter is the kernel informing the policy that a folio was actually evicted. This distinction exists for the following two reasons. A folio can be evicted in circumvention of the ``normal'' eviction path if, for example, the file containing it is deleted. Conversely, in rare cases, proposing a folio for eviction does not guarantee that it will be evicted (\eg the folio is in active use by the kernel).

We use eBPF's \texttt{struct\_ops} feature in order to minimize the verifier changes needed to add new eBPF hooks. \texttt{struct\_ops} was designed to allow kernel subsystems to expose modular interfaces to eBPF components, and has already been used for TCP congestion control algorithms, FUSE eBPF filesystems, handling HID driver quirks, and \texttt{sched\_ext}~\cite{sched_ext,fuse-bpf,bpf-extensible-network,hid-bpf}. \texttt{struct\_ops} programs are loaded into the kernel like any other eBPF program.
Using \texttt{struct\_ops} also makes it much easier to extend \name and add new hooks. For example, we implemented an extension to \name that added a page cache admission filter with only 15 additional lines of verifier-related code.

\begin{figure}
    \centering
    \begin{lstlisting}[language=C]
// Policy function hooks
struct cachebpf_ops {
    s32  (*policy_init)(struct mem_cgroup *memcg);
    // Propose folios to evict
    void (*evict_folios)(struct eviction_ctx *ctx, struct mem_cgroup *memcg);
    void (*folio_added)(struct folio *folio);
    void (*folio_accessed)(struct folio *folio);
    // Folio was removed: clean up metadata
    void (*folio_removed)(struct folio *folio);
    char name[CACHEBPF_OPS_NAME_LEN];
};

struct eviction_ctx {
    u64 nr_candidates_requested; /* Input  */
    u64 nr_candidates_proposed;  /* Output */
    struct folio *candidates[32];
};
    \end{lstlisting}
    \caption{\texttt{struct\_ops} for \name and eviction context.}
    \label{fig:cache-ext-ops}
\end{figure}

\subsubsection{Eviction Lists}
\label{subsubsec:design_eviction_lists}

Eviction algorithms are implemented on a wide range of data structures. 
Nevertheless, we observe that many of these policies can be implemented either exactly or approximately using linked lists, where the policy iterates over one or more lists and evicts items based on a calculated per-item score. For example, the ``classic'' eviction policies, (\eg LRU, MRU) are all based on lists, with items inserted or evicted from the head or tail of a list. Similarly, families of policies like ARC~\cite{arc} or segmented LRU~\cite{segmented} can be implemented using multiple variable-sized lists, where items are inserted into any list or moved between lists. Even recent ``state-of-the-art'' policies, such as LHD, S3-FIFO, or LRB either store data directly in a list~\cite{s3-fifo,sieve}, or simply select a sample of objects and evict the ones with the lowest \emph{score}~\cite{lhd,lrb}. 

In order to facilitate an interface flexible enough for all these policies, \name is built around an \emph{eviction list API}, a simple interface for policies to construct and manipulate a set of variable-sized linked lists. Each node in the list corresponds to a single folio, and stores a pointer to that folio, rather than the folio itself. Importantly, the actual folios are still stored and maintained by the default kernel page cache implementation, in order to minimize changes to the kernel.

This API is implemented as a set of eBPF \texttt{kfuncs} (in-kernel functions that are exposed to eBPF programs) and is shown in Table~\ref{table:cache-ext-eviction-lists-api}.\footnote{The actual functions have a ``\name'' prefix to prevent name collisions, but we omit it for brevity.}
For example, \texttt{init()} will typically call \texttt{list\_create()} to create a new eviction list, and \texttt{folio\_added()} will call \texttt{list\_add()} to add the folio to a list. Newly-created lists are added to a ``registry'', an internal per-policy hash table which maps between the list IDs (exposed to eBPF) and the lists themselves. Notably, these lists are \emph{indexed} -- that is, given a folio pointer, the APIs can directly obtain that folio's list node. This property is necessary for operations such as deletion from the list, and is facilitated using a per-policy hash-table which maps from folios to list nodes. We discuss the use of this hash table further in \S\ref{subsec:design_safe_memory_referencing}.

\begin{table}[t!]
\centering
\small
\begin{tabular}{|p{0.95\columnwidth}|}
\hline
\multicolumn{1}{|c|}{\textbf{Eviction list API}} \\
\hline
\texttt{u64 list\_create(struct mem\_cgroup *memcg)}\\
\hline
\texttt{int list\_add(u64 list, struct folio *f, bool tail)}\\
\hline
\texttt{int list\_move(u64 list, struct folio *f, bool tail)}\\

\hline
\texttt{int list\_del(struct folio *f)}\\
\hline
\texttt{int list\_iterate(struct mem\_cgroup *memcg}\\
\texttt{\quad u64 list,}\\
\texttt{\quad s64(*iter\_fn)(int id, struct folio *f),}\\
\texttt{\quad struct iter\_opts *opts,} \\
\texttt{\quad struct eviction\_ctx *ctx)} \\
\hline
\end{tabular}
\caption{\name eviction list API.}
\label{table:cache-ext-eviction-lists-api}
\end{table}

\subsubsection{Eviction Candidate Interface}
\label{subsubsec:eviction-interface}
To facilitate eviction, policy functions iterate over their eviction lists in order to determine which folios to evict. Note that policies do not directly evict folios -- rather, they propose \emph{eviction candidates} to the kernel, which checks if the folios are indeed valid eviction targets (\ie not pinned or in other use by the kernel) and evicts them if so.

The eBPF framework currently does not provide a clean way to iterate over the eviction lists, so \name provides a new iteration \texttt{kfunc} which allows policy functions to specify how to iterate over a list and make decisions for each node. Specifically, \texttt{list\_iterate()} takes a list to iterate over, an options struct, an eviction context, and a callback function. The callback function, which is also an eBPF program, is called on each node, and the policy decides whether to keep or evict that folio. Folios chosen for eviction are added to the \texttt{candidates} array in the \texttt{eviction\_ctx} struct. The options struct specifies how the interface should treat evaluated folios. For example, they can be left in place, moved to the tail of the list, or moved to a different list. This enables implementing policies that make use of multiple lists and require balancing the lists, such as S3-FIFO or ARC. We also provide a ``batch scoring'' mode for this interface, where the callback function is used to compute \emph{scores} for $N$ folios, with the $C$ folios with the lowest score chosen for eviction. This mode can be used for policies such as LFU.

In order to ensure correct verification of our callback functions, we added $\sim$80 lines to the eBPF verifier to ``register'' our iteration interfaces, on top of eBPF's existing support for callback functions. Additionally, to protect against eBPF program misbehavior, this interface performs the requisite bounds-checking and enforces loop termination.

\subsubsection{eBPF limitations}
\label{subsubsec:ebpf-limitations}

We ran into a number of challenges when implementing \name's eviction list API. eBPF maps, the standard way to maintain state in eBPF programs, do not provide interfaces that both store items in a specified order while also providing random access, both of which are necessary in order to implement eviction properly. Specifically, eBPF provides maps such as \texttt{BPF\_MAP\_TYPE\_QUEUE} and \texttt{BPF\_MAP\_TYPE\_STACK}, which provide \texttt{pop()} and \texttt{push()} operations, but do not allow deleting or accessing elements from the ``middle'' of the map. Conversely, \texttt{BPF\_MAP\_TYPE\_HASH} provides random access, but no method to easily maintain an ordering of elements (\eg MRU order). A notable exception is \texttt{BPF\_MAP\_TYPE\_LRU\_HASH}, which provides both an LRU structure and random-access, but is too deeply tied to its specific algorithm for our purposes~\cite{bpf_map_type_hash}. This necessitated the development of a custom data structure for \name.

While eBPF has started to introduce experimental support for custom data structures and more complex locking in eBPF, this support is not yet mature enough for our use case~\cite{kflex,bpf-arena,red-black-tree-bpf}. As such, we designed our list API to be managed by the kernel and exposed to eBPF via \texttt{kfuncs}. Additionally, in order to avoid concurrency issues and verifier limitations around locking, the provided API is concurrency-safe and makes use of locks under the hood, in the kernel implementations. As eBPF matures, new features could further reduce overhead and provide even more flexibility for eBPF policies.

\subsubsection{Example: LFU Policy}
\label{subsubsec:design_mru_example}

\begin{figure}
\begin{lstlisting}[language=C,basicstyle=\ttfamily\small]
u64 lfu_list;
int lfu_policy_init(struct mem_cgroup *cg) {
    lfu_list = list_create(cg);
    return 0;
}
void lfu_folio_added(struct folio *folio) {
    u64 freq = 1;
    list_add(lfu_list, folio, true); // Add to tail
    bpf_map_update_elem(&freq_map, &folio, &freq);
}
void lfu_folio_accessed(struct folio *folio) {
    u64 *freq = bpf_map_lookup_elem(&freq_map, &folio);
    __sync_fetch_and_add(freq, 1); // Increment freq
}
long score_lfu(int id, struct folio *folio) {
    return bpf_map_lookup_elem(&freq_map, &folio);
}
void lfu_evict_folios(struct eviction_ctx *ctx, struct mem_cgroup *cg) {
    struct iter_opts opts = { /* Set scoring mode */ };
    list_iterate(cg, lfu_list, score_lfu, &opts, ctx);
}
void lfu_folio_removed(struct folio *folio) {
    bpf_map_delete_elem(&freq_map, &folio);
}
\end{lstlisting}
\caption{Simplified LFU implementation with \name.}
\label{figure:example-lfu}
\end{figure}

To get a better sense of how \name's policy functions can be used to implement custom policies, we walk through implementing a simple eviction policy, LFU, using \name. LFU evicts the least-frequently accessed item in the list, which requires storing additional metadata. Our LFU implementation uses a single list and an eBPF map to store folio access frequencies. It approximates LFU using \name's batch scoring mode to select the $C$ (\eg 32) least-frequently accessed folios out of $N$ (\eg 512) folios.

A simplified version of the policy is shown in Figure~\ref{figure:example-lfu}. When the policy is loaded and \texttt{lfu\_policy\_init()} is called, we create a new eviction list. When a folio is added, \texttt{lfu\_folio\_added()} adds the folio to the tail of the list using \texttt{list\_add()} and initializes its frequency to 1 in the \texttt{freq\_map} eBPF map (not shown). When a folio is accessed, we increment its frequency. When eviction is triggered, \texttt{lfu\_evict\_folios()} calls \texttt{list\_iterate()}, which calls the \texttt{score\_lfu()} callback function on $N$ nodes in the list. The score function returns the frequency of each folio as its score. \name then selects the $C$ folios with the lowest scores as eviction candidates. When a folio is evicted by the kernel, \texttt{lfu\_folio\_removed()} is called, and the folio's metadata is removed from the map. Additionally, it is not necessary to remove the folio from the list on eviction, as this is taken care of by \name. We discuss this point further in \S\ref{subsec:design_safe_memory_referencing}.

\subsection{Isolation}
\label{subsec:design_isolation}

We now tackle the third challenge from \S\ref{sec:challenges}: how to allow applications to deploy their own policy functions without interfering with other applications' policies, while preserving the sharing property of the page cache, whereby applications can avoid having to load duplicate pages into memory. 

We make the observation that implementing policies within a cgroup can address this challenge. %
This is due to the fact that within a cgroup, processes have the same custom eviction policy, and different cgroups running on the same server can each use their own eviction policy. In addition, deploying policy functions per-cgroup fits the common pattern of deploying modern applications via containers, which isolate each application in its own memory cgroup. Note that processes from cgroup A can still access page cache memory managed by cgroup B, and benefit from accessing shared data. %

To support per-cgroup policies, we extend eBPF's \texttt{struct\_ops} functionality to support cgroup-specific \texttt{struct\_ops} (currently, it only supports system-wide policies). This involved adding a cgroup identifier (in the form of a file descriptor) to the kernel's \texttt{struct\_ops} loading interface, along with corresponding libbpf interfaces in userspace. %

\subsection{Memory Safety}
\label{subsec:design_safe_memory_referencing}

We must ensure that \name does not allow unsafe memory accesses when interacting with folio pointers (challenge 4 from \S\ref{sec:challenges}). Specifically, \name must ensure that eBPF programs return valid pointers to the kernel (\ie in the eviction candidate interface). Otherwise, a malicious eBPF program could return invalid values, leading to memory corruption or a kernel crash. Frameworks like \texttt{sched\_ext} solve this in part by using PIDs as identifiers for processes in userspace dispatch. However, folios do not have analogous easily-obtainable unique identifiers, so we resort to using folio pointers.

In order to validate these pointers, we implement a registry of ``valid folios'' in the system. When a folio enters the page cache, \name adds it to the registry. When a folio is evicted, it is removed from the registry. When a \name eviction proposes a set of folio eviction candidates, the kernel uses the registry to verify that each candidate is indeed a valid folio before proceeding with eviction. This registry is implemented as a hash table with a per-bucket lock, which also stores a folio's list node (as described in \S\ref{subsubsec:design_eviction_lists}), which maps from folio pointer to list node. We find that this design incurs minimal overhead, which we evaluate in \S\ref{subsec:eval-overhead}. Future developments in eBPF may make it easier to keep track of ``trusted'' pointers, potentially allowing us to remove this check and further reduce overhead.

We also protect against adversarial behavior by providing a fallback for eviction. For example, if the kernel asks a faulty policy to evict 10 folios, but it only proposes 5 candidates, the kernel will fall back to its default policy and evict additional folios. Similarly, when a folio is evicted, the kernel ensures that it is removed from any eviction list it is present in, in order to release memory resources and minimize stale references lying around. Similar fallbacks are present in other frameworks, such as \texttt{sched\_ext}, which implements a watchdog that forcibly removes misbehaving policies.

\subsection{Kernel Implementation Complexity}
\label{subsec:kernel-complexity}

Implementing \name required adding $\sim$2000 lines to the kernel. Only a fraction of these lines touched core kernel code: 210 lines in the page cache (most of which are the new eBPF hooks), 80 lines in the verifier (supporting our callback functions), and 80 lines in cgroup code. Additionally, implementing per-cgroup \texttt{struct\_ops} required 220 lines in the kernel and 75 lines in libbpf. The remaining lines implemented pure \name functionality: 750 lines for \name's eviction list \texttt{kfuncs}, and 580 lines to implement registry operations and register \name with the verifier.

\section{Evaluation}
\label{sec:evaluation}
We aim to answer the following questions:
\begin{denseenum}
\item[{\bf Q1:}] Is \name flexible enough to implement a variety of eviction policies? Can \name policies improve application performance with low developer effort? (\S\ref{subsec:eval-custom-eviction-policies})
\item[{\bf Q2:}] Can different applications use different policies without interfering with each other? (\S\ref{subsec:eval-isolation})
\item[{\bf Q3:}] What is the overhead of \name? (\S\ref{subsec:eval-overhead})
\end{denseenum}

\paragraph{System configuration.}
We conduct our experiments on Cloudlab~\cite{cloudlab} c6525-25g machines, with a 16-core AMD Rome CPU, 128GB of memory and a 480GB SSD drive. We use CPU-pinning and disable SMT, swap, and address space randomization to make our results more reproducible. We also drop the page cache before each test. We run Ubuntu 22.04 with Linux v6.6.8 as the kernel.

\subsection{Case Studies: Custom Policies (Q1)}
\label{subsec:eval-custom-eviction-policies}

In this section, we describe several case studies on how applications can utilize \name to achieve better performance. First, we show how \name can be used to implement a wide range of eviction policies, tailored to different applications: from simple ``classic'' policies (MRU and LFU) to state-of-the-art policies such as LHD~\cite{lhd}, which uses conditional probabilities to model different page features (\eg age, frequency), and S3-FIFO~\cite{s3-fifo}.
 We then explore how an application can make its eviction policy \emph{aware} of application-specific information, such as assigning different priorities to specific types of requests. %

\subsubsection{Most-Recently Used (MRU)}
Scan-like workloads do not perform well under LRU-like policies when the scan length is larger than the size of the LRU list. For example, one such scan-heavy workload is searching through files in a codebase. Consider a developer working on a large codebase, such as the Linux kernel, and continuously searching for certain terms. %
In such a scenario, an LRU-like policy would evict the files that were least-recently searched, but those are precisely the files that are required at the start of the next search. While readahead can help mitigate this issue for single-file scans by prefetching sequential blocks, it cannot predict which blocks to fetch across files.

\begin{figure}[t!]
  \centering
  \includegraphics[width=0.33\textwidth]{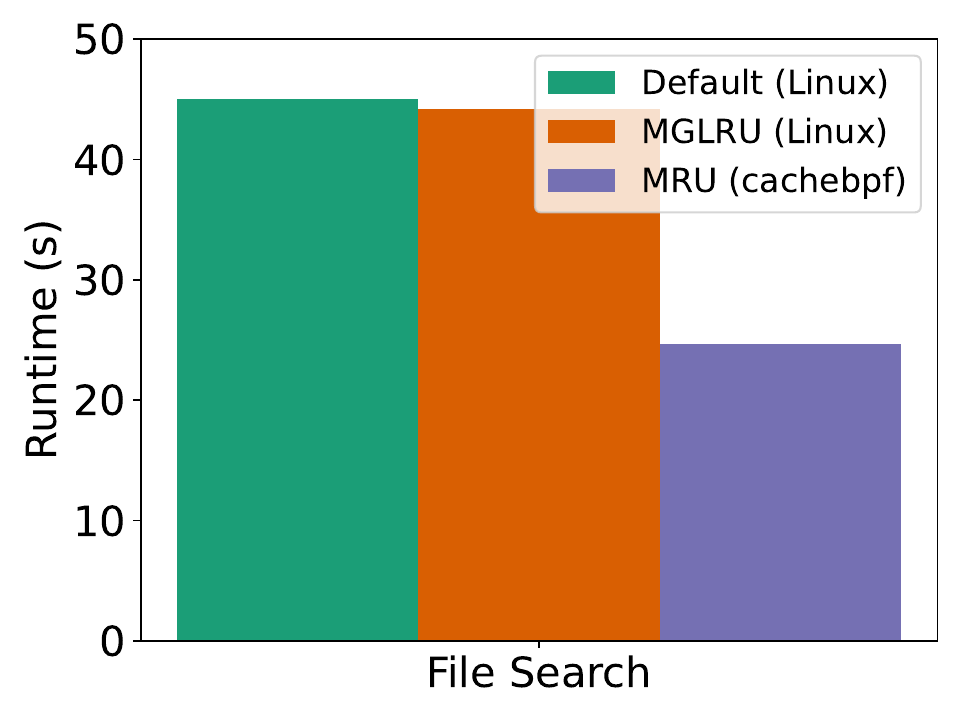}
  \caption{File search workload results (MRU policy).}
  \label{fig:eval-mru-filesearch}
\end{figure}

\begin{figure*}[!t]
    \centering
    \begin{subfigure}[t]{0.37\textwidth}
        \centering
        \includegraphics[width=\columnwidth]{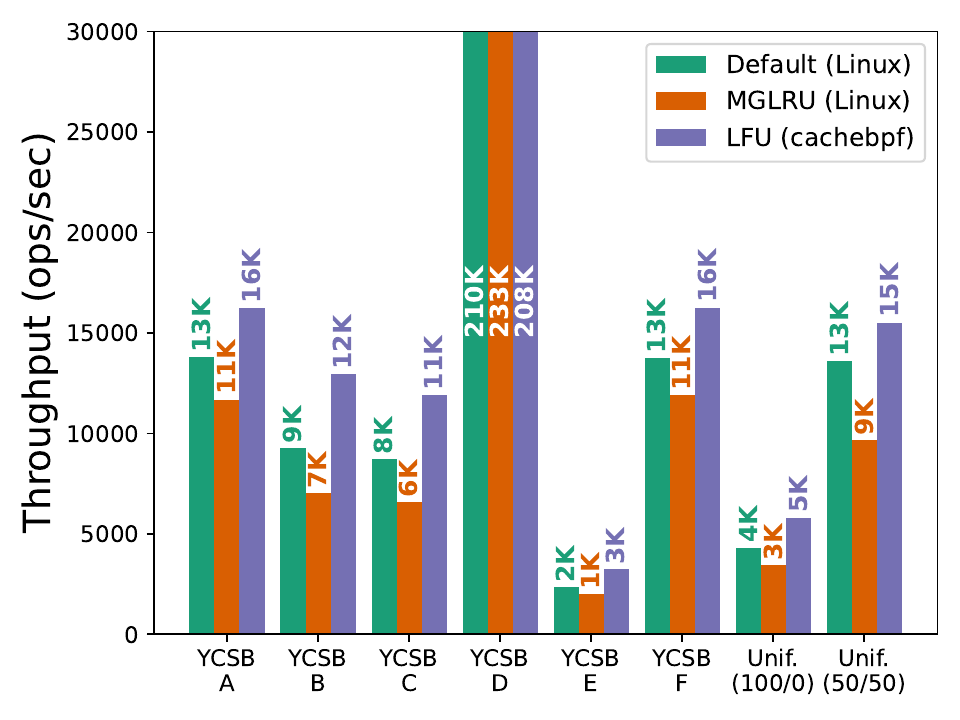}
        \caption{Throughput.}
        \label{subfig:ycsb-results-throughput}
    \end{subfigure}
    ~~~~~~~~~~~~~~~~~~~~~
    \begin{subfigure}[t]{0.37\textwidth}
        \centering
        \includegraphics[width=\columnwidth]{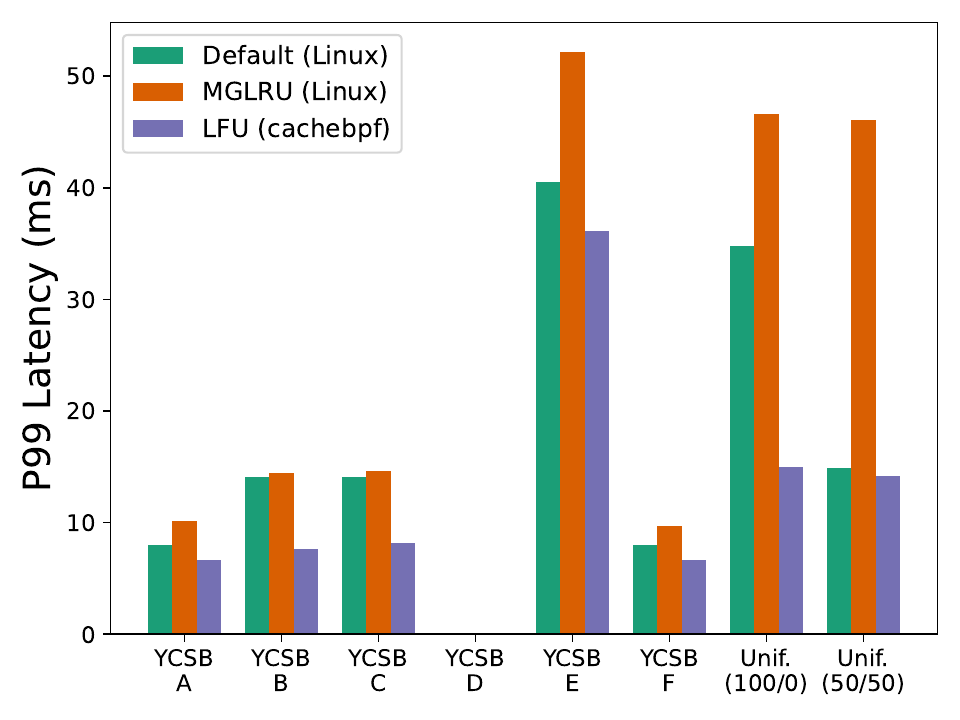}
        \caption{P99 latency.}
        \label{subfig:ycsb-results-latency}
    \end{subfigure}
    \caption{YCSB workload results (LFU policy).}
    \vspace{-0.5em}
    \label{fig:eval-lfu-ycsb}
\end{figure*}

We use \name to develop an MRU eviction policy for this use case. In contrast to LRU, MRU will evict the folios that were most recently searched, and will be used again furthest in the future.
In order to facilitate this, our policy adds folios to the head of the list on insertion, and moves them back to the head on access. No metadata is required, as nodes are stored in the correct order in the eviction list. In a simplified version of the policy, when eviction is triggered, the first 32 nodes in the list are selected as eviction candidates using \name's iterate interface.
However, if the policy decides to evict folios right after they were added to the page cache, they may still be in use by the kernel to service the I/O request. This would lead to the kernel refusing to evict the folios and resorting to the fallback path to evict folios. Therefore, we skip a small fixed number of folios when iterating the eviction list before proposing eviction candidates.

\vspace{-1em}
\paragraph{Evaluation.} To evaluate the policy, we construct a file search workload that searches the Linux kernel codebase (v6.6), using the multi-threaded \emph{ripgrep} CLI tool~\cite{ripgrep}. More specifically, we perform 10 searches within a 1GiB cgroup, which is roughly 70\% the size of the codebase (excluding Git history). We compare the \name MRU policy with the default Linux kernel policy as well as the kernel's experimental MGLRU policy. The results in Figure~\ref{fig:eval-mru-filesearch} show that \name is almost 2$\times$ faster than both baseline and MGLRU, since both policies suffer from the scan ``pathology'' of LRU.

\subsubsection{Least-Frequently Used (LFU)}
\label{subsubsec:LFU}
An additional disadvantage of LRU-like algorithms is that they only take into account a single feature (recency) when making eviction decisions.
However, other features, such as access frequency, may be better suited for certain workloads, especially skewed workloads where the access distribution is static or slow-changing. One such workload is the popular YCSB benchmark, made to model cloud OLTP applications, in which the probability of each key being requested is drawn independently from a static distribution (by default, Zipfian).

We use \name to implement an LFU policy, which takes access frequency into account when evicting folios by evicting those with the lowest access frequency among the eviction candidates. Our LFU policy is an \emph{approximate} LFU policy, as it does not evict the global least-frequently used folio, but only the least-frequently ones among the current batch of folios considered for eviction. An exact policy would either yield higher overhead or require more complex data structures, which eBPF does not yet support. %
We described the implementation of our LFU policy in \S\ref{subsubsec:design_mru_example}.

\vspace{-1em}
\paragraph{Evaluation.} We evaluate our LFU policy by running LevelDB~\cite{leveldb}, a popular key-value store, on the YCSB (Zipfian) workloads, as well as against uniform and uniform-read-write workloads. We compare this custom policy against both the default and MGLRU Linux policies, using a 100GiB database with a 10GiB cgroup. 
Our results in Figure~\ref{fig:eval-lfu-ycsb} show that \name's LFU policy outperforms both the default and MGLRU, for all the YCSB Zipfian workloads and the uniform workloads, except for YCSB D, which only uses the latest key-value pairs and as such is cached entirely in-memory. \name achieves up to 37\% better throughput than the default Linux policy, and interestingly, it outperforms MGLRU by an even greater margin. We also measure the P99 read latency, for which \name beats the default policy by up to 55\%. Note that YCSB D's tail latency barely registers in the figure due to its lack of disk accesses. %
We also evaluated the YCSB workload with our other policies, but LFU outperformed those policies as well, so we omit those results.

\vspace{0.5em}
\noindent\fbox{%
    \parbox{0.97\columnwidth}{%
        \textbf{Takeaway 1:}
        \name can significantly improve application performance even with simple policies (\eg MRU, LFU) that match the application's access patterns. 
    }%
}

\subsubsection{S3-FIFO}
S3-FIFO~\cite{s3-fifo} is a recent caching policy designed for key-value caches, which uses three FIFO queues to quickly remove ``one-hit wonders'' (keys that are accessed only once). It has been shown to yield significant throughput gains on a number of workloads. We implement S3-FIFO using \name and evaluate it on Twitter production cache traces below.

S3-FIFO uses a main FIFO and a small FIFO to hold $\sim$90\% and 10\% of the objects, respectively. Upon insertion, objects are typically added to the small FIFO. It uses a ghost FIFO to track recently-evicted objects, in order to promote them quickly to the main FIFO on readmission. The small FIFO is used to filter out short-lived objects, while objects that are accessed more often are promoted from the small FIFO to the main FIFO. The access frequency of the objects is tracked, but is capped at a maximum of 3, in order to ensure that a burst of accesses does not prevent objects from being evicted.

We implement the main and small FIFOs as two eviction lists, and the ghost FIFO as a \texttt{BPF\_MAP\_TYPE\_LRU\_HASH} map.
The map will then automatically remove entries from the ghost FIFO in LRU order when it hits capacity.
When a folio is evicted, we create a ghost entry using a pointer to its \texttt{struct address\_space} (which represents a file's contents), along with the folio's offset in the file, as the key. Note that we cannot use folio pointers as the key, as they are not persistent across evictions. While we cannot implement the ghost FIFO as an eviction list (as they operate on \emph{valid} folios), it is more performant and simpler to use an existing eBPF map. We find that the combination of existing eBPF features and \name is sufficiently flexible to implement complex eviction policies.

On folio addition, we set its access frequency in an eBPF map, and update it on access. We use eviction candidate requests to evict folios, but also to maintain the 90\%-10\% ratio between the main and small lists. If the small list is over-represented, we evict from it. We use \name's eviction iteration interface: if a folio's access frequency is greater than 1, we move it to the tail of the main list, balancing the lists. Otherwise, we propose the folio for eviction, and move it to the tail of the small list so that it isn't considered again before it is evicted. When evicting from the main list, we use the iteration interface to find folios with access frequency of 0. %
If we cannot find enough of those, we search for folios with access frequency of 1, and then 2, and so on. All folios that are considered for eviction have their access frequency decremented and are moved to the tail of the main list.

\subsubsection{Least Hit Density (LHD)}
LHD is a relatively sophisticated eviction policy that uses conditional probabilities to predict which objects are most likely to be accessed in the future~\cite{lhd}. LHD uses a metric called \emph{hit density} in order to determine which objects should be evicted, along with a \emph{dynamic ranking} approach which allows it to automatically tune its eviction policy over time. We implement LHD using \name, based on the implementation in libcachesim~\cite{twittertraces,s3-fifo,sieve}.

Our implementation only uses one eviction list. However, folios are divided into \emph{classes} partially based on when they were last accessed and their age at that time. Each class stores its own statistics (\eg hits, evictions, hit densities) for different folio ages. Folios are not explicitly ``owned'' by classes -- instead, they use metadata from classes based on which class they most closely correspond to at a given time. When a folio is added, we store its metadata, such as its last access time and age at that time, in an eBPF map. That metadata is updated on folio access, and removed when the folio is evicted.

Folios are selected as eviction candidates based on their hit density (or more accurately, the hit density of the class and age they most closely correspond to). LHD iterates over the list and selects the folios with the lowest hit density as eviction candidates. While this process is fairly straightforward, it is enabled by accurate computation of hit densities over time. LHD requires periodically ``reconfiguring'' its hit densities and other statistics in order to ensure that its probability distributions are accurate and aged appropriately over time using an exponentially weighted moving average (EWMA).

This reconfiguration process needs to run every $N$ folio admissions or insertions (where $N$ is a relatively large number -- \eg $2^{20}$). However, reconfiguration is a relatively expensive process, requiring iterating over all of the policy's metadata and adjusting it. In order to avoid performing this operations in the page cache's insertion or access hot paths, we use an eBPF ring buffer to notify userspace that reconfiguration needs to take place. Userspace then calls an eBPF program of type \texttt{BPF\_PROG\_TYPE\_SYSCALL}, which allows running an eBPF program without attaching it to a specific hook. This program then performs the required reconfiguration, including computing updated hit densities, and scaling or compressing distributions as necessary. We use atomic operations to ensure that the page cache can continue using these values, albeit with some potential inaccuracy, which we permit for the sake of performance. While we could implement this reconfiguration step in userspace, doing so would have required numerous syscalls to interact with eBPF maps, and atomic updates would not have been possible. Additionally, we note that in a standard LHD policy, hit densities and other parameters are stored as floating-point values. However, eBPF does not support floating-point operations, so we resort to scaling values by a large constant in order to approximate such calculations.

\vspace{-1em}
\paragraph{Evaluation.}
For many real-world workloads, it may not be obvious in advance which policy works best for a given workload. \name makes experimentation easy, allowing developers to implement a set of policies and \emph{empirically} choose the best one for each workload.

\begin{figure}[t!]
    \centering
    \includegraphics[width=0.4\textwidth]{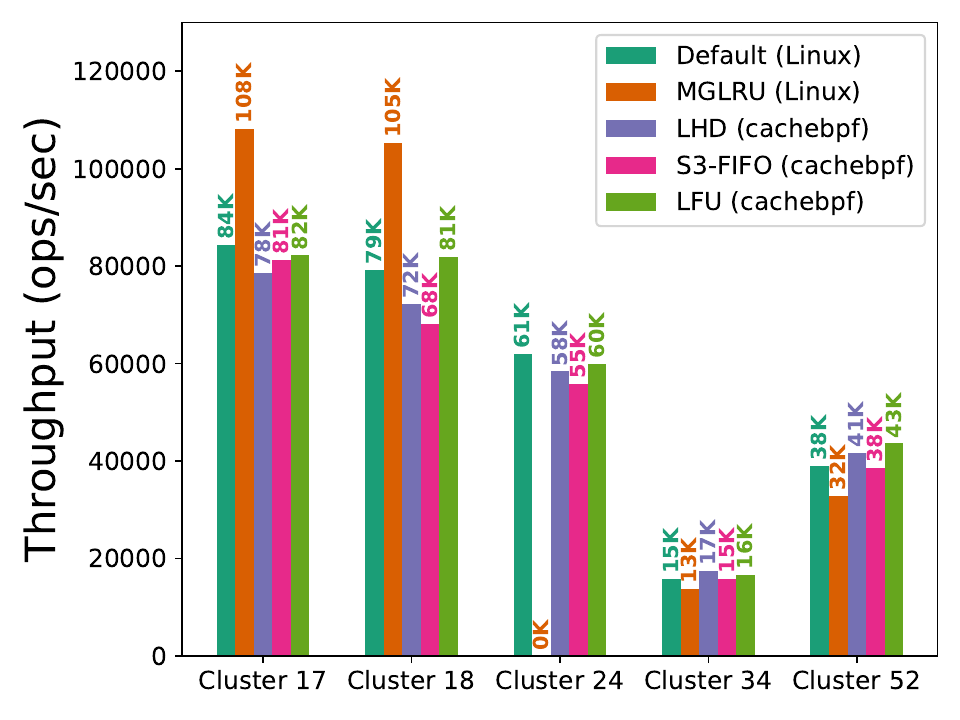}
    \caption{Twitter workload results (LHD, S3-FIFO, and LFU policies) using LevelDB. No one policy performs best across the different clusters.}
    \label{fig:eval-twitter}
\end{figure}

We evaluate our LHD, LFU, and S3-FIFO policies on production traces taken from the Twitter cache workloads~\cite{twittertraces}. The workloads divide the traces by cluster ID. We compare these policies to Linux's default and MGLRU policies. Each cluster was evaluated with a cgroup size set to 10\% of the cluster's data size using LevelDB. As shown in Figure~\ref{fig:eval-twitter}, we find that, in general, there is no single policy that is best for all workloads. While LHD beats the default and MGLRU policies by 13\% and 30\%, respectively, on cluster 34, and LFU beats them by 13\% and 34\% on cluster 52, MGLRU dominates on clusters 17 and 18. In cluster 24, the default policy is best, while MGLRU consistently resulted in out-of-memory errors, hence its 0 throughput result. Meanwhile, S3-FIFO beats or matches the baseline on clusters 34 and 52, but does not outperform the other \name policies.

\begin{figure}[t!]
    \centering
    \includegraphics[width=0.4\textwidth]{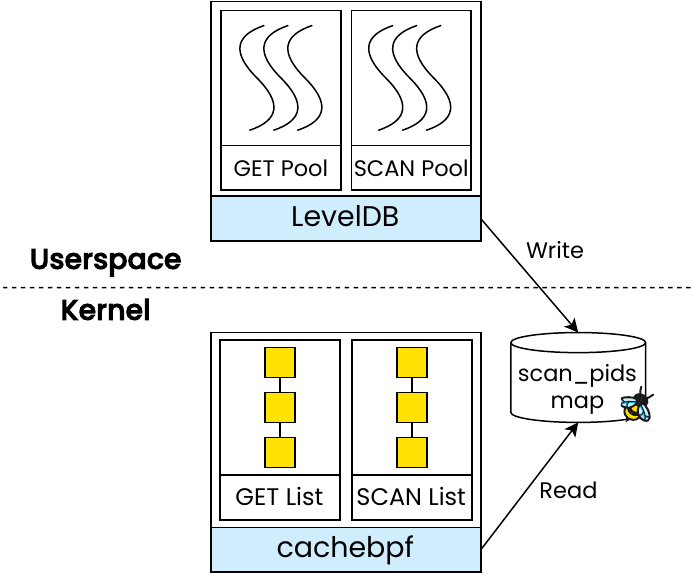}
    \caption{Overview of GET-SCAN policy implementation.}
    \label{fig:mixed-get-scan-system}
\end{figure}

\vspace{0.5em}
\noindent\fbox{%
    \parbox{0.97\columnwidth}{%
        \textbf{Takeaway 2:}
        There is no one-size-fits-all policy that performs best for all workloads. Customization and experimentation are necessary to maximize performance.
    }%
}

\subsubsection{Application-Informed Policy (GET-SCAN)}
\label{sec:app-informed}
In addition to enabling the implementation of a variety of general eviction algorithms, \name enables applications to use eviction algorithm tailored to their design. In other words, the eviction algorithm can be made \emph{aware of application-level abstractions} and uses this information to make better decisions.
To illustrate the value of application-informed policies, consider the case of heterogeneous queries in databases. For example, a database serving financial transactions could see many small queries for individual payments, while also performing slower scan-like queries in the background to conduct fraud detection, reconciliation, and other business processes. While these scan-like queries are important, they typically have more relaxed service-level objectives. However, these large scan-like requests can ``pollute'' the page cache and degrade the performance of the smaller requests, as generic eviction algorithms struggle to isolate the folios used by these requests. Ideally, the page cache should prioritize the small requests over the large ones in the presence of memory pressure. Using \name, we can build an application-informed policy that fulfills this requirement.

To simulate the application we describe above, we run LevelDB with a mixed SCAN/GET workload. This workload is highly skewed and is composed of 99.95\% GET requests with a small amount of SCANs (0.05\%). We use a separate thread-pool for SCAN requests to avoid head-of-line blocking at the scheduling level, as per prior work~\cite{syrup}, with a disjoint set of threads handling GET requests. While the workload exhibits good cache locality for GETs, it has poor locality for SCANs, which span a large number of folios and exhibit high reuse distance. The existing kernel eviction policy cannot handle this scenario well, leading to cache pollution due to a large number of SCAN folios.

Using \name, we design a policy that is aware of the different request types. We observe that a folio accessed by a SCAN should not be worth the same as a folio accessed by a GET. To implement prioritization, the policy uses two eviction lists: one for folios inserted by GET requests, and the other for those inserted by SCAN requests. When loading the policy, the application initializes an eBPF map with the PIDs of the SCAN threads. When a folio is inserted, the policy checks whether the PID of the current task is present in the map to determine which eviction list to add the folio to. Each eviction list independently maintains an approximate LFU policy, as described in \S\ref{subsubsec:LFU}. When the kernel requests eviction candidates, the policy prioritizes evicting folios from the SCAN list. Figure~\ref{fig:mixed-get-scan-system} illustrates this policy.

\begin{figure}[t]
    \centering
    \begin{subfigure}[b]{0.49\columnwidth}
        \centering
        \includegraphics[width=\textwidth]{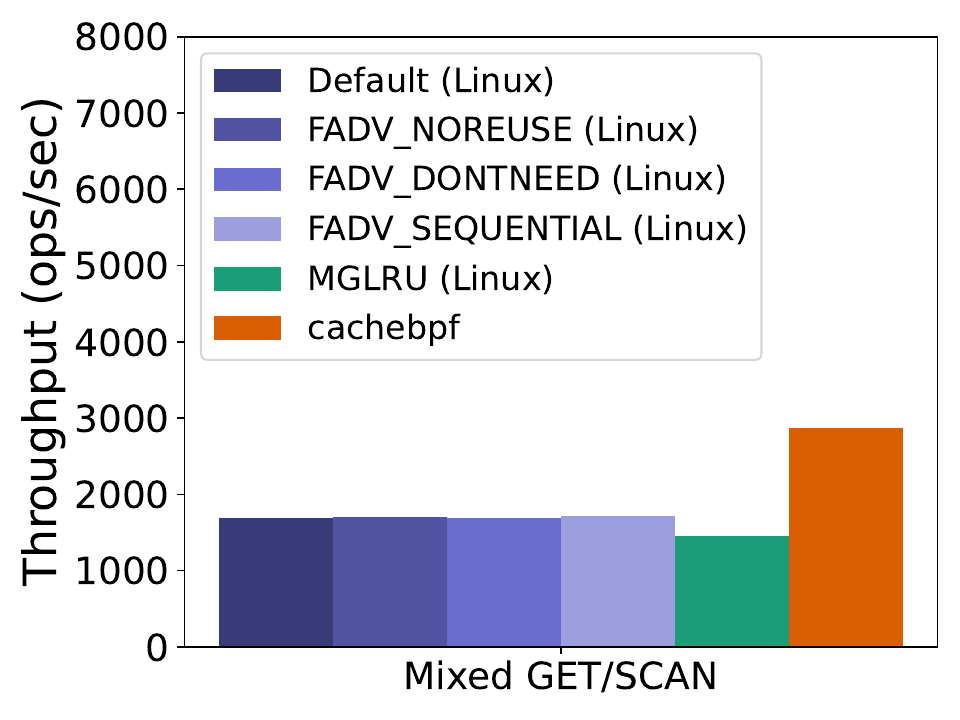}
        \caption{GET throughput.}
        \label{subfig:mixed-get-scan-results-throughput}
    \end{subfigure}
    \hfill
    \begin{subfigure}[b]{0.49\columnwidth}
        \centering
        \includegraphics[width=\textwidth]{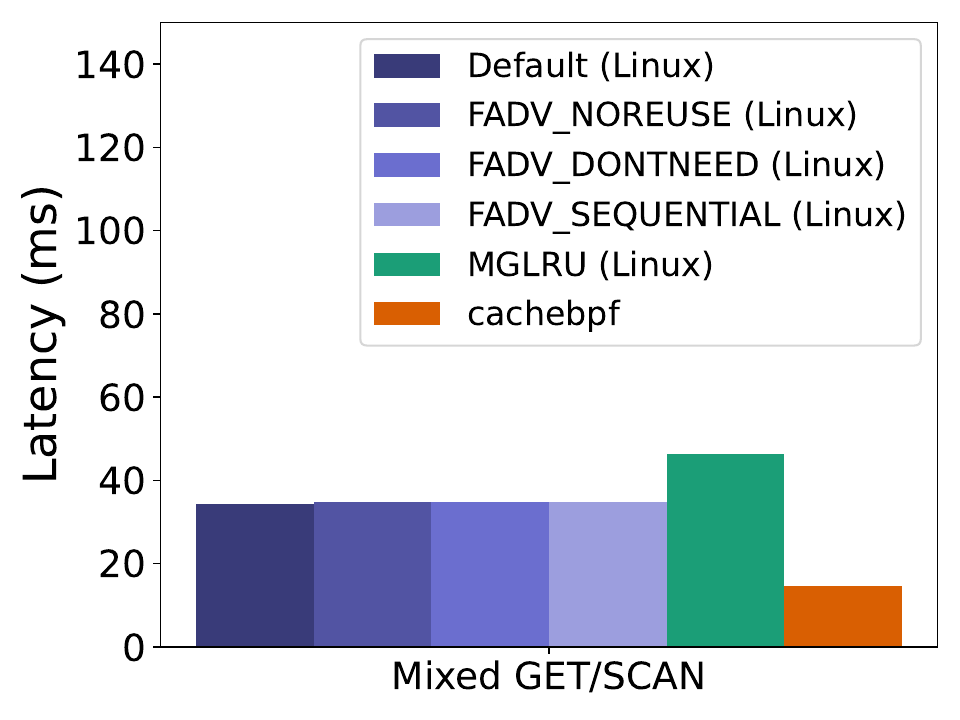}
        \caption{GET P99 latency.}
        \label{subfig:mixed-get-scan-results-latency}
    \end{subfigure}
    \caption{Mixed GET-SCAN workload results.}
    \label{fig:mixed-get-scan-results}
\end{figure}

\vspace{-1em}
\paragraph{Evaluation.} To evaluate the policy, we compare against Linux's default and MGLRU policies, and various \texttt{fadvise()} options: \texttt{FADV\_DONTNEED}, \texttt{FADV\_NOREUSE} and \texttt{FADV\_SEQUENTIAL} (on top of the default policy). We apply these \texttt{fadvise()} options to files used by SCAN requests, in order to inform the kernel in advance that we plan to read the files sequentially or only once (\texttt{SEQUENTIAL} and \texttt{NOREUSE}) or that we no longer need the folios after their use (\texttt{DONTNEED}).
As shown in Figure~\ref{fig:mixed-get-scan-results}, \name's application-informed policy achieves 1.70$\times$ the throughput and 57\% lower P99 latency for GET requests, while SCAN requests experience an 18\% throughput decrease. In addition, the \texttt{fadvise()} options do not help much, demonstrating the inadequacy of existing kernel page cache interfaces compared to \name. MGLRU performs even worse than the default LRU.

\vspace{0.5em}
\noindent\fbox{%
    \parbox{0.97\columnwidth}{%
        \textbf{Takeaway 3:}
        Even very simple application-aware eviction policies can significantly improve performance.
    }%
}

\noindent\fbox{%
    \parbox{0.97\columnwidth}{%
        \textbf{Takeaway 4:}
        Existing Linux page cache customization interfaces are ineffective.
    }%
}

\vspace{-1em}
\subsubsection{Implementation Complexity}

Table~\ref{tab:loc-per-policy} shows the lines of eBPF and userspace loader code necessary to implement each of the aforementioned policies, along with a simple FIFO policy.
The policies are all implemented in at most a few hundred lines of code, a much smaller amount than would be necessary to implement them within the kernel (or even in userspace).
We find that \name reduces the complexity of developing new policies, by using the pre-defined list and policy function abstractions, as well as by relying on the kernel's existing page cache to actually store the folios. In addition, developer experience and velocity is greatly improved, since eBPF prevents kernel crashes and many types of bugs, enabling developers to focus on the policy logic. Thus, \name allows developers to accelerate their applications with a relatively modest amount of effort.

Additionally, we plan to open source all of our policies, allowing developers to easily try them with their applications, lowering the barrier to entry for using \name.

\begin{table}
\centering
\footnotesize
\begin{tabular}{lrr}
\toprule
Policy & eBPF LoC & Userspace LoC \\
\midrule
FIFO & 56 & 118 \\
MRU & 101 & 87 \\
LFU & 221 & 107 \\
S3-FIFO & 287 & 139 \\
GET-SCAN & 324 & 107 \\
LHD & 366 & 152 \\
\bottomrule
\end{tabular}
\caption{Lines of eBPF and userspace loader code in \name policies.}
\label{tab:loc-per-policy}
\end{table}

\begin{figure}[t!]
    \centering
    \includegraphics[width=0.38\textwidth]{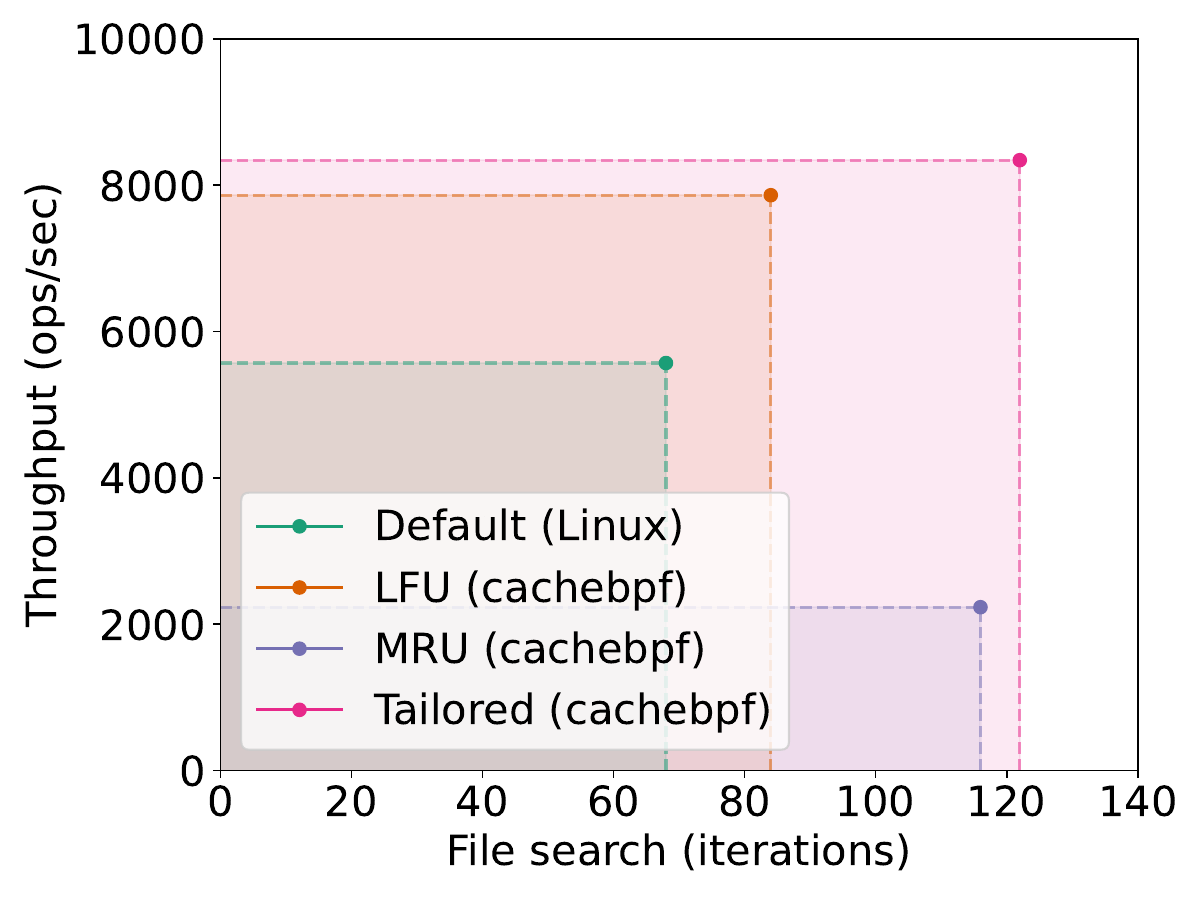}
    \caption{Using \name, multiple applications can run different eviction policies, yielding better performance for all.}
    \label{fig:eval-isolation}
\end{figure}

\subsection{Isolation (Q2)}
\label{subsec:eval-isolation}

The Linux page cache already provides a measure of isolation by giving each cgroup its own set of LRU lists. \name takes advantage of this design by enabling each cgroup to have its own custom policy. We demonstrate that this is a useful capability by simulating and comparing against ``global'' policies, as opposed to \name's per-cgroup policies. We create two cgroups, one running a YCSB C workload with LevelDB, and the other running a file search workload with ripgrep. The YCSB cgroup is allocated 10GiB and the file search cgroup is allocated 1GiB. We run these workloads under four configurations: both cgroups using the default policy, both using LFU, both using MRU, and a ``tailored'' setup: YCSB with LFU and file search with MRU.

\vspace{-1em}
\paragraph{Evaluation.} Figure~\ref{fig:eval-isolation} shows that the tailored setup beats the other three configurations, yielding 49.8\% and 79.4\% improvements for YCSB and file search, respectively, over the baseline configuration. While the other two \name configurations provide performance improvements for the workloads corresponding to their policy, they can significantly degrade the performance of the other workload, demonstrating that global policies are indeed not a viable solution. Note that YCSB improves further in the tailored setup compared to the LFU configuration (and vice versa for file search compared to the MRU configuration). This is due to improved caching of the workloads yielding reduced disk contention. Additionally, the file search workload improves in the LFU configuration for the same reason.

\vspace{0.5em}
\noindent\fbox{%
    \parbox{0.97\columnwidth}{%
        \textbf{Takeaway 5:}
        Using \name with per-cgroup policies allows for fine-grained control and improved performance.
    }%
}

\subsection{Memory and CPU Overhead (Q3)}
\label{subsec:eval-overhead}

The advent of faster and larger storage devices means that the page cache (and \name) must be able to handle millions of events per second. We run a number of micro-benchmarks to investigate \name's memory and CPU overhead.

\vspace{-1em}
\subsubsection{Memory Overhead}
\name's primary memory usage is the valid folios registry hash table (\S\ref{subsec:design_safe_memory_referencing}).
In the worst case, we set up the hash table with as many buckets as there are 4KiB pages in the cgroup (based on its configured size). Each bucket requires 16 bytes to store the hash table's internal list pointers. Thus, the memory overhead for an empty registry is:
$$\frac{\left(\frac{\text{cgroup\_size}}{\text{page\_size}}\right) \times 16}{\text{cgroup\_size}} = 0.4\%$$
Each filled entry in the hash table uses 32 bytes for the \name list node. The full registry memory overhead is:
$$\frac{\left(\frac{\text{cgroup\_size}}{\text{page\_size}}\right) \times (16 + 32)}{\text{cgroup\_size}} = 1.2\%$$
Therefore, the memory overhead for \name's registry is between 0.4\%-1.2\% of a policy's cgroup's memory. We believe that this overhead could be significantly reduced with recent improvements to eBPF's handling of kernel objects, allowing eBPF to directly ensure that some pointers are trusted.

\subsubsection{CPU Overhead}
To measure the CPU overhead, we run the fio micro-benchmark~\cite{fio} with 8 threads on a \emph{randread} workload and record the IOPS and CPU usage. We do this for both the default Linux policy and a \name no-op policy, meaning that it uses the default eviction policy while still maintaining \name data structures. Then, we calculate and compare the CPU usage per I/O operation (measured in $\mu$CPUs, \ie one-millionth of a CPU). Table~\ref{tab:eval-cpu-overhead} shows that the CPU overhead of \name is at most 1.7\%.

\begin{table}
\centering
\footnotesize{}
\begin{tabular}{lrrr}
\toprule
Cgroup Size & Baseline &     \name & Overhead (\%) \\
\midrule
      5 GiB &   234.80 &    236.51 &        0.72\% \\
     10 GiB &   217.48 &    221.14 &        1.66\% \\
     30 GiB &   197.67 &    198.01 &        0.17\% \\
\bottomrule
\end{tabular}
\caption{\name $\mu$CPU usage per I/O operation using fio.}
\label{tab:eval-cpu-overhead}
\end{table}

\vspace{0.5em}
\noindent\fbox{%
    \parbox{0.97\columnwidth}{%
        \textbf{Takeaway 6:}
        \name incurs relatively low CPU and memory overhead, while improving performance.
    }%
}

\section{Related Work}
\label{sec:related}

There have been three predominant approaches to allow applications to customize the page cache. 
The first approach, which was explored in the 80's and 90's, was to design clean-slate extensible kernels~\cite{SPIN,VINO,exokernel-original,mach}, which allow applications to customize kernel interfaces and policies. For example, VINO~\cite{VINO,VINO-2} and SPIN~\cite{SPIN} %
allow applications to customize buffer cache eviction, admission, and prefetching policies. These OS designs never achieved widespread use, even though some of their underlying ideas have become relevant again with the adoption of eBPF, which enables extensibility within monolithic kernels like Linux or Windows.

The second approach, introduced in the 90's, is to design customizable file systems, which allow applications to customize the page cache. ACFS~\cite{kai-li-caching,kai-li-caching2} is an application-controlled file system which enables customizing caching and prefetching. The XN~\cite{exokernel} libOS file system enables running a userspace-level file system within the exokernel OS, which can be fully customized. More recent work in this vein is Bento~\cite{bento}, which allows custom file systems written in Rust to be installed in the kernel, without disrupting applications. None of these approaches would work with existing Linux or legacy file systems. %

The third approach is for applications to simply implement their own userspace cache, with the option of bypassing the OS page cache with direct I/O. There are many examples of data systems that implement a userspace cache~\cite{rocksdb,wisckey,splinterdb}. TriCache~\cite{tricache} is a recent framework that helps applications customize their own userspace caches. Nonetheless, many popular data systems still rely on the page cache, sometimes in conjunction with userspace caches~\cite{rocksdb,postgres,wiredtiger,symbiosis,combining-caches}. %

There has been more recent work on customizing memory management policies using eBPF, such as huge page placement, page fault handling, and page table designs~\cite{prog-mem-bpf,custom-page-fault,ebpf-mm}. Most relevant to \name, FetchBPF allows customizing Linux's memory prefetching policy, and could easily be integrated into \name as an additional hook~\cite{fetchbpf}. P2Cache is conceptually similar to \name, but only allows LRU or MRU ordering and changes the global page cache policy~\cite{P2Cache}. Additionally, the P2Cache paper is a work-in-progress workshop paper, is closed-source, and does not contain many details about its design, implementation, or evaluation.

\section{Conclusion}

This work explores the design of a new eBPF framework to implement custom eviction policies in the kernel, enabling applications to choose a policy according to their needs and making the latest caching research accessible to the kernel. We believe there are significant future research challenges in this area, such as exploring ML-based eviction algorithms and integrating more parts of the page cache into the \name framework (\eg writeback and prefetching). Furthermore, we are aware of efforts in the eBPF verifier to support more complex data structures and believe that \name could benefit from these efforts.

\label{lastpage}

\section{Acknowledgments}
We would like to thank Kostis Kaffes, Tanvir Ahmed Khan, and Yuhong Zhong for their feedback.
We also thank the CloudLab team for their help in supporting our experiments.
This work was supported by IBM, and NSF awards CNS-2143868 and CNS-2106530.
Tal Zussman was supported by NSF award DGE-2036197.
Ioannis Zarkadas is an Onassis Foundation scholar.

\bibliographystyle{plain}
\bibliography{database}

\clearpage

\end{document}